# An Actinide-boost Star Discovered in the Gaia-Sausage-Enceladus

Yangming Lin[1,2], Haining Li[1], Ruizheng Jiang[1,2], Wako Aoki[3,4], Satoshi Honda[5], Zhenyu He[6,7], Ruizhi Zhang[1], Zhuohan Li[1,2], and Gang Zhao[1,2]
[1] CAS Key Laboratory of Optical Astronomy, National Astronomical Observatories, Chinese Academy of Sciences, Beijing 100101, People's Republic of China; lhn@nao.cas.cn
[2] School of Astronomy and Space Science, University of Chinese Academy of Sciences, No.19(A) Yuquan Road, Shijingshan District, Beijing 100049, People's Republic of China
[3] National Astronomical Observatory of Japan, 2-21-1 Osawa, Mitaka, Tokyo 181-8588, Japan
[4] Astronomical Science Program, The Graduate University for Advanced Studies, SOKENDAI, 2-21-1 Osawa, Mitaka, Tokyo 181-8588, Japan
[5] Nishi-Harima Astronomical Observatory, Center for Astronomy, University of Hyogo, 407-2, Nishigaichi, Sayo-cho, Sayo, Hyogo 679-5313, Japan
[6] School of Physics, Peng Huanwu Collaborative Center for Research and Education, and International Research Center for Big-Bang Cosmology and Element Genesis, Beihang University, Beijing 100191, People's Republic of China
[7] ShenYuan Honors College, Beihang University, Beijing 100191, People's Republic of China
Received 2025 January 19; revised 2025 March 25; accepted 2025 April 1; published 2025 May 6

## Abstract

We report the discovery of an actinide-boost, very metal-poor ([Fe/H] = −2.38), r-process-enhanced ([Eu/Fe] = 0.80) star, LAMOST J0804+5740, within the Gaia-Sausage-Enceladus (GSE). Based on the high-resolution ($R \sim 36{,}000$ and 60,000) and high signal-to-noise ratio spectra obtained with the High Dispersion Spectrograph on the Subaru Telescope, the abundances of 48 species are determined. Its $\log \epsilon \, (\mathrm{Th/Eu}) = -0.22$ establishes it as the first confirmed actinide-boost star within the GSE. Comparative analysis of its abundance pattern with theoretical r-process models reveals that the magnetorotationally driven jet supernova r-process model with $\hat{L}\nu = 0.2$ provides the best fit and successfully reproduces the actinide-boost signature. Kinematic analysis of actinide-boost stars reveals that approximately two-thirds of them are classified as ex situ stars, suggesting that actinide-boost stars are more likely to originate from accreted dwarf galaxies. As the first actinide-boost star identified within the GSE, J0804+5740 will provide valuable insights into r-process nucleosynthesis in accreted dwarf galaxies like the GSE, especially on the production of the heaviest elements.

*Unified Astronomy Thesaurus concepts:* Chemically peculiar giant stars (1201); Stellar abundances (1577); R-process (1324); Galactic archaeology (2178); Milky Way stellar halo (1060)

*Materials only available in the* online version of record: *machine-readable table*

## 1. Introduction

The rapid neutron-capture process (r-process) and slow neutron-capture process (s-process) are primary nucleosynthesis mechanisms responsible for producing elements heavier than iron in the Universe. The r-process is believed to occur under extreme physical conditions, yet its astrophysical site remains a highly debated topic. The gravitational-wave event GW170817 and its associated electromagnetic kilonova counterpart AT2017gfo (B. P. Abbott et al. 2017) have provided compelling evidence that neutron star mergers (NSMs) can produce heavy elements (A. Frebel & A. P. Ji 2023). However, the NSM scenarios face challenges in explaining the observed [Eu/Fe] and [Fe/H] relationship in very metal-poor (VMP) stars in the Milky Way due to their long merging timescales (C. Kobayashi et al. 2023). Other potential astrophysical sites for r-process include magnetorotationally driven jet supernovae (MRDSNe; C. Winteler et al. 2012; N. Nishimura et al. 2015, 2017) and collapsar jets (A. I. MacFadyen & S. E. Woosley 1999; D. M. Siegel et al. 2019; Z. He et al. 2024). However, these astrophysical sites lack direct observational evidence. Determining the astrophysical sites and yield of the r-process, as well as the frequency of r-process events, remains a persistent goal.

VMP ([Fe/H] < −2.0) stars (T. C. Beers & N. Christlieb 2005) are considered among the oldest objects in the Universe, with their surfaces preserving the chemical signatures of their formation. For r-process-enhanced VMP stars in the halo of the Milky Way, the r-process elements on their surfaces are believed to originate from a single r-process event. These stars are classified according to their [Eu/Fe] and [Ba/Eu]: r-I stars have $0.3 \leqslant [\mathrm{Eu/Fe}] \leqslant 1.0$ and [Ba/Eu] < 0.0, r-II stars have [Eu/Fe] > 1.0 and [Ba/Eu] < 0.0 (T. C. Beers & N. Christlieb 2005), while r-III stars have [Eu/Fe] > 2.0 and [Ba/Eu] < 0.0 (M. Cain et al. 2020). CS 22892-052 was the first extreme r-process-enhanced star discovered (C. Sneden et al. 1994), with [Eu/Fe] exceeding that of the Sun by more than 10 times. After nearly 30 yr of unremitting efforts (T. C. Beers et al. 1992; P. S. Barklem et al. 2005; N. Christlieb et al. 2008; E. M. Holmbeck et al. 2020; H. Li et al. 2022), approximately 480 r-I stars, 165 r-II stars, and 2 r-III stars have been discovered (M. Cain et al. 2020; H. Li et al. 2022; D. Shank et al. 2023; I. U. Roederer et al. 2024).

Numerous studies have established that the abundance patterns from the range of Ba to Pt in r-process-enhanced stars with varying metallicity and Eu abundance are consistent and align with the solar r-process residual. However, this homogeneity breaks down in the region of light neutron-capture elements (Z < 56) and the heaviest neutron-capture elements (Z = 90, 92). Approximately one-third of r-process-enhanced stars exhibit $\log \epsilon \, (\mathrm{Th/Eu})$ that are roughly 0.4 dex higher than those of other r-process-enhanced stars (L. Mashonkina et al. 2014;

C. M. Sakari et al. 2018). This phenomenon is known as actinide boost (V. Hill et al. 2002), but the underlying cause of the overall high-actinide abundances in these stars remains unidentified.

In this Letter, we report the discovery and present a detailed abundance analysis covering the full *r*-process pattern of an actinide-boost star within the Gaia-Sausage-Enceladus (GSE), J0804+5740. In Section 2, we introduce the high-resolution spectral observations and the methodology for determining the atmospheric parameters. Detailed information on the derived element abundances is provided in Section 3. In Section 4, we discuss the abundance pattern of J0804+5740, the potential connections with the fission of transuranic nuclei, as well as the dynamical signatures of *r*-process-enhanced metal-poor stars with measurable Th abundances. Our summary and conclusions are presented in Section 5.

## 2. Observations and Data Analysis

J0804+5740 was classified as an *r*-II VMP star due to its elevated [Eu/Fe], based on high-resolution ($R \sim 36{,}000$) spectra covering a wavelength range of 4030–6800 Å (H. Li et al. 2022), obtained with the Subaru Telescope/High Dispersion Spectrograph (HDS; K. Noguchi et al. 2002). Additionally, based on its highly radial/eccentric orbit characteristics, it was identified as a member of the GSE substructure (R. Zhang et al. 2024). To further investigate its properties, high-resolution ($R \sim 60{,}000$) blue-end spectra were obtained, with a total exposure time of 2400 s. By employing the StdBc setup with a 0″.6 slit and 2 × 2 CCD binning, a signal-to-noise ratio (S/N) of 142 per pixel at 4500 Å was achieved, covering the range from 3550 to 5210 Å.

The effective temperature ($T_{\rm eff}$) and surface gravity ($\log g$) of J0804+5740 are determined photometrically, based on the method described in H. Li et al. (2022). We prefer to use photometry and parallax to determine the $T_{\rm eff}$ and $\log g$ since Fe I lines are significantly affected by non–local thermodynamic equilibrium effects in metal-poor giants. We adopt the color–$T_{\rm eff}$–[Fe/H] relationship for $T_{\rm eff}$ from I. Ramírez & J. Meléndez (2005). Specifically, we use the $(V - K_s)$ color index, which is less sensitive to variations in surface gravity and metallicity (A. Alonso et al. 1999). The $V$ and $K_s$ magnitudes are sourced from the APASS and Two Micron All Sky Survey catalogs, with extinction estimated from G. M. Green et al. (2018). We use reliable parallax measurements ($\sigma_{\rm parallax}/{\rm parallax} = 0.06$) from Gaia DR3, along with the parallax zero-point correction provided by Y. Huang et al. (2021). To calculate the star's surface gravity, the following formula is employed:

$$\log g = \log g_\odot + \log \frac{M}{M_\odot} + 4 \log \frac{T_{\rm eff}}{T_{\rm eff\odot}} + 0.4(M_{\rm bol} - M_{\rm bol\odot}), \quad (1)$$

where the stellar mass ($M$) is assumed to be 0.8 $M_\odot$ (H. Li et al. 2022). Additionally, the following reference values for the Sun are adopted: $\log g_\odot = 4.438$, $\log T_{\rm eff\odot} = 3.7617$, and $M_{\rm bol\odot} = 4.75$ (S. P. Shah et al. 2024). Furthermore, we apply the bolometric correction in the $V$ band using the relationship described by A. Alonso et al. (1999).

Due to the interdependence of atmospheric parameters, we determine the final values through an iterative process. Specifically, the determination of microscopic turbulence involves minimizing the trend in the abundance of Fe I lines and their reduced equivalent widths ($\log {\rm EW}/\lambda$). Additionally, the abundance obtained from Fe II lines is adopted to determine [Fe/H]. The resulting atmospheric parameter values obtained in this study are summarized in Table 1. These values are consistent with those reported in H. Li et al. (2022). It is worth noting that the [Fe/H] in H. Li et al. (2022) was determined using the abundance obtained from Fe I lines.

**Table 1**
Basic Parameters of J0804+5740

| Parameter | Value | References |
|---|---|---|
| R.A. (hh:mm:ss) | 08:04:52.8 | Gaia Collaboration et al. (2023) |
| Decl. (deg:arcmin:arcsec) | 57:40:19.4 | Gaia Collaboration et al. (2023) |
| $\ell$ (deg) | 159.86 | Gaia Collaboration et al. (2023) |
| $b$ (deg) | 32.36 | Gaia Collaboration et al. (2023) |
| $V$ mag | 11.510 ± 0.097 | U. Munari et al. (2014) |
| $K_s$ mag | 8.859 ± 0.016 | R. M. Cutri et al. (2003) |
| $E(B-V)$ (mag) | 0.051 ± 0.011 | G. M. Green et al. (2018) |
| $RV_{\rm helio}$ (km s$^{-1}$) | −318.36 ± 0.15 | This study |
| $RV_{\rm helio}$ (km s$^{-1}$) | −318.08 ± 0.20 | Gaia Collaboration et al. (2023) |
| $RV_{\rm helio}$ (km s$^{-1}$) | −320.18 ± 3.15 | A. L. Luo et al. (2022) |
| Parallax (mas) | 0.284 ± 0.017 | Gaia Collaboration et al. (2023) |
| Distance (kpc) | 3.054 ± 0.181 | F. Anders et al. (2022) |
| $r_{\rm apo}$ (kpc) | 21.80 ± 0.12 | This study |
| $r_{\rm peri}$ (kpc) | 0.53 ± 0.20 | This study |
| $Z_{\rm max}$ (kpc) | 12.68 ± 2.65 | This study |
| ecc | 0.953 ± 0.018 | This study |
| $E$ (10$^3$ km$^2$ s$^{-2}$) | −131.74 ± 0.20 | This study |
| $J_r$ (kpc km s$^{-1}$) | 1555.22 ± 57.48 | This study |
| $J_\phi$ (kpc km s$^{-1}$) | 226.39 ± 82.81 | This study |
| $J_z$ (kpc km s$^{-1}$) | 202.00 ± 19.86 | This study |
| $T_{\rm eff}$ (K) | 4585 ± 85 | This study |
| $T_{\rm eff}$ (K) | 4556 ± 93 | H. Li et al. (2022) |
| $\log g$ (cgs) | 1.41 ± 0.07 | This study |
| $\log g$ (cgs) | 1.45 ± 0.07 | H. Li et al. (2022) |
| $\xi$ (km s$^{-1}$) | 1.82 ± 0.02 | This study |
| $\xi$ (km s$^{-1}$) | 1.94 ± 0.05 | H. Li et al. (2022) |
| [Fe/H] | −2.38 ± 0.09 | This study |
| [Fe/H] | −2.47 ± 0.09 | H. Li et al. (2022) |

## 3. Element Abundance

We measure equivalent widths (EWs) for over 600 lines and derive the abundances for 48 species. The atomic parameters of the lines used to derive the abundances, along with the EW measurements, are presented in Table A1. Table 2 lists the final abundances, along with their associated uncertainties, the number of measured lines, and the method to derive the abundance. Solar abundances are adopted from M. Asplund et al. (2009). The species abundances are derived from EW measurements or spectral synthesis. Figure 1 shows examples of spectral syntheses for key lines. We adopt the 1D plane-parallel hydrostatic model atmosphere from the ATLAS NEWODF grid by F. Castelli & R. L. Kurucz (2003) to analyze the abundance of all species, assuming local thermodynamic equilibrium and no convective overshoot. The EWs of the various lines are measured with the Tool for Automatic Measurement of Equivalent Width (TAME, W. Kang &

Table 2
Element Abundances of J0804+5740

| Species | $\log \epsilon_\odot$ | $\log \epsilon$ | [X/H] | [X/Fe] | $\sigma_{stat}$ | $\sigma_{sys}$ | $\sigma_{\log \epsilon}$ | $\sigma_{[X/Fe]}$ | N | Method[a] |
|---|---|---|---|---|---|---|---|---|---|---|
| C (CH) | 8.43 | 5.63 | −2.80 | −0.42 | 0.09 | 0.10 | 0.13 | 0.16 | 1 | S |
| N (CN) | 7.83 | 5.90 | −1.93 | 0.45 | 0.09 | 0.18 | 0.20 | 0.22 | 1 | S |
| O I | 8.69 | 7.15 | −1.54 | 0.84 | 0.09 | 0.06 | 0.11 | 0.14 | 1 | S |
| Na I | 6.24 | 4.08 | −2.16 | 0.22 | 0.10 | 0.15 | 0.18 | 0.20 | 2 | S |
| Mg I | 7.60 | 5.57 | −2.03 | 0.35 | 0.04 | 0.11 | 0.12 | 0.15 | 10 | E |
| Al I | 6.45 | 3.70 | −2.75 | −0.37 | 0.09 | 0.15 | 0.18 | 0.20 | 1 | S |
| Si I | 7.51 | 5.50 | −2.01 | 0.37 | 0.06 | 0.12 | 0.14 | 0.16 | 2 | E |
| Ca I | 6.34 | 4.21 | −2.13 | 0.25 | 0.02 | 0.08 | 0.08 | 0.12 | 32 | E |
| Sc II | 3.15 | 0.92 | −2.23 | 0.15 | 0.03 | 0.03 | 0.04 | 0.10 | 15 | S |
| Ti I | 4.95 | 2.63 | −2.32 | 0.06 | 0.02 | 0.14 | 0.14 | 0.17 | 29 | E |
| Ti II | 4.95 | 2.99 | −1.96 | 0.42 | 0.02 | 0.03 | 0.04 | 0.10 | 40 | E |
| V I | 3.93 | 1.32 | −2.61 | −0.23 | 0.05 | 0.15 | 0.16 | 0.18 | 3 | S |
| V II | 3.93 | 1.63 | −2.30 | 0.08 | 0.04 | 0.03 | 0.05 | 0.10 | 8 | S |
| Cr I | 5.64 | 2.89 | −2.75 | −0.37 | 0.02 | 0.14 | 0.14 | 0.17 | 14 | E |
| Cr II | 5.64 | 3.40 | −2.24 | 0.14 | 0.05 | 0.03 | 0.06 | 0.11 | 5 | E |
| Mn I | 5.43 | 2.45 | −2.98 | −0.60 | 0.04 | 0.10 | 0.11 | 0.14 | 11 | S |
| Fe I | 7.50 | 5.05 | −2.45 | −0.07 | 0.01 | 0.12 | 0.12 | 0.15 | 128 | E |
| Fe II | 7.50 | 5.12 | −2.38 | 0.00 | 0.02 | 0.02 | 0.09 | 0.13 | 18 | E |
| Co I | 4.99 | 2.40 | −2.59 | −0.21 | 0.04 | 0.16 | 0.17 | 0.19 | 9 | S |
| Ni I | 6.22 | 3.67 | −2.55 | −0.17 | 0.02 | 0.10 | 0.10 | 0.14 | 19 | E |
| Cu I | 4.19 | 1.08 | −3.11 | −0.73 | 0.09 | 0.13 | 0.16 | 0.18 | 1 | S |
| Zn I | 4.56 | 2.33 | −2.23 | 0.15 | 0.06 | 0.02 | 0.07 | 0.11 | 2 | E |
| Sr II | 2.87 | 0.94 | −1.93 | 0.45 | 0.06 | 0.07 | 0.10 | 0.13 | 2 | S |
| Y II | 2.21 | −0.03 | −2.24 | 0.14 | 0.03 | 0.06 | 0.07 | 0.11 | 14 | E |
| Zr II | 2.58 | 0.78 | −1.80 | 0.58 | 0.03 | 0.05 | 0.06 | 0.11 | 21 | E |
| Nb II | 1.46 | <−0.01 | <−1.47 | <0.91 | ... | ... | ... | ... | 1 | S |
| Mo I | 1.88 | 0.03 | −1.85 | 0.53 | 0.09 | 0.16 | 0.18 | 0.21 | 1 | S |
| Ru I | 1.75 | 0.21 | −1.54 | 0.84 | 0.06 | 0.16 | 0.17 | 0.19 | 4 | S |
| Rh I | 0.91 | −0.33 | −1.24 | 1.14 | 0.12 | 0.17 | 0.21 | 0.23 | 2 | S |
| Ba II | 2.18 | 0.18 | −2.00 | 0.38 | 0.06 | 0.07 | 0.10 | 0.13 | 5 | S |
| La II | 1.10 | −0.73 | −1.83 | 0.55 | 0.01 | 0.06 | 0.06 | 0.11 | 18 | S |
| Ce II | 1.58 | −0.37 | −1.95 | 0.43 | 0.02 | 0.06 | 0.06 | 0.11 | 39 | E |
| Pr II | 0.72 | −1.00 | −1.72 | 0.66 | 0.02 | 0.07 | 0.07 | 0.11 | 6 | S |
| Nd II | 1.42 | −0.32 | −1.74 | 0.64 | 0.02 | 0.07 | 0.07 | 0.12 | 50 | E |
| Sm II | 0.96 | −0.71 | −1.67 | 0.71 | 0.02 | 0.05 | 0.06 | 0.11 | 31 | E |
| Eu II | 0.52 | −1.06 | −1.58 | 0.80 | 0.02 | 0.07 | 0.08 | 0.12 | 5 | S |
| Gd II | 1.07 | −0.54 | −1.61 | 0.77 | 0.04 | 0.05 | 0.06 | 0.11 | 16 | E |
| Tb II | 0.30 | −1.29 | −1.59 | 0.79 | 0.06 | 0.07 | 0.09 | 0.13 | 2 | E |
| Dy II | 1.10 | −0.48 | −1.58 | 0.80 | 0.04 | 0.05 | 0.07 | 0.11 | 13 | E |
| Ho II | 0.48 | −1.19 | −1.67 | 0.71 | 0.05 | 0.06 | 0.08 | 0.12 | 3 | S |
| Er II | 0.92 | −0.71 | −1.63 | 0.75 | 0.04 | 0.06 | 0.08 | 0.12 | 8 | E |
| Tm II | 0.10 | −1.49 | −1.59 | 0.79 | 0.08 | 0.05 | 0.09 | 0.13 | 3 | E |
| Yb II | 0.84 | −0.81 | −1.65 | 0.73 | 0.09 | 0.06 | 0.11 | 0.14 | 1 | S |
| Lu II | 0.10 | −1.45 | −1.55 | 0.83 | 0.09 | 0.03 | 0.09 | 0.13 | 1 | S |
| Hf II | 0.85 | −0.77 | −1.62 | 0.76 | 0.09 | 0.05 | 0.11 | 0.14 | 1 | S |
| Os I | 1.40 | −0.19 | −1.59 | 0.79 | 0.17 | 0.17 | 0.24 | 0.26 | 2 | S |
| Ir I | 1.38 | −0.14 | −1.52 | 0.86 | 0.09 | 0.12 | 0.15 | 0.18 | 1 | S |
| Pb I | 1.75 | 0.15 | −1.60 | 0.78 | 0.09 | 0.13 | 0.16 | 0.18 | 1 | S |
| Th II | 0.02 | −1.28 | −1.30 | 1.08 | 0.09 | 0.08 | 0.12 | 0.15 | 1 | S |
| U II | −0.54 | <−1.87 | <−1.33 | <1.05 | ... | ... | ... | ... | 1 | S |

**Note.**
[a] The method used to derive the abundance is indicated as follows: "E" denotes that the abundance is derived from EWs, while "S" signifies that the abundance is derived from spectral synthesis.

S.-G. Lee 2012), which fits theoretical Gaussian/Voigt profiles to normalized observed spectra. Additionally, we use an updated version of the abundance analysis code MOOG (C. A. Sneden 1973), treating Rayleigh scattering as coherent, isotropic scattering (J. S. Sobeck et al. 2011). We compile line lists from various literature sources (R. Cayrel et al. 2004; W. Y. Cui et al. 2013; L. Mashonkina et al. 2014; V. Hill et al. 2017; M. Gull et al. 2018; I. U. Roederer et al. 2018; H. Li et al. 2022; S. P. Shah et al. 2024) and updated their atomic parameters from VALD (T. Ryabchikova et al. 2015). Molecular and blended atomic line lists for spectral synthesis are generated from Linemake (V. M. Placco et al. 2021), an open-source atomic and molecular line list generation tool.

The uncertainties in elemental abundances are determined by two primary components: uncertainties arising from measurements and those associated with atmospheric parameters, denoted as $\sigma_{\rm stat}$ and $\sigma_{\rm sys}$, respectively, and presented in columns (6) and (7) of Table 2. The $\sigma_{\rm stat}$ mainly originates from the uncertainties in the measurements of EWs, the oscillator strength of lines, and line blending. It can be estimated from $\frac{\sigma_X}{\sqrt{N}}$, where $\sigma_X$ represents the standard deviation of the abundance of species $X$ derived from $N$ lines. For cases where the number of adopted lines is fewer than four, $\sigma_{\rm stat}$ is taken as the larger value of $\frac{\sigma_X}{\sqrt{N}}$ and $\frac{\sigma_{\rm Fe\,II}}{\sqrt{N}}$. The $\sigma_{\rm sys}$ primarily arises from the uncertainties in atmospheric parameters. By individually varying the $T_{\rm eff}$, $\log g$, $\xi$, and [Fe/H] by $1\sigma$ and calculating the square root of the quadratic sum corresponding to the change in abundance, we estimate $\sigma_{\rm sys}$. Notably, neutral transitions of molybdenum (Mo I), ruthenium (Ru I), rhodium (Rh I), and osmium (Os I) exhibit the highest sensitivity to these parameter changes, with $\sigma_{\rm sys}$ in their derived abundances exceeding 0.15 dex. The total error ($\sigma_{\log\epsilon}$) is calculated as the square root of the sum in quadrature of $\sigma_{\rm stat}$ and $\sigma_{\rm sys}$ and is listed in column (8) of Table 2.

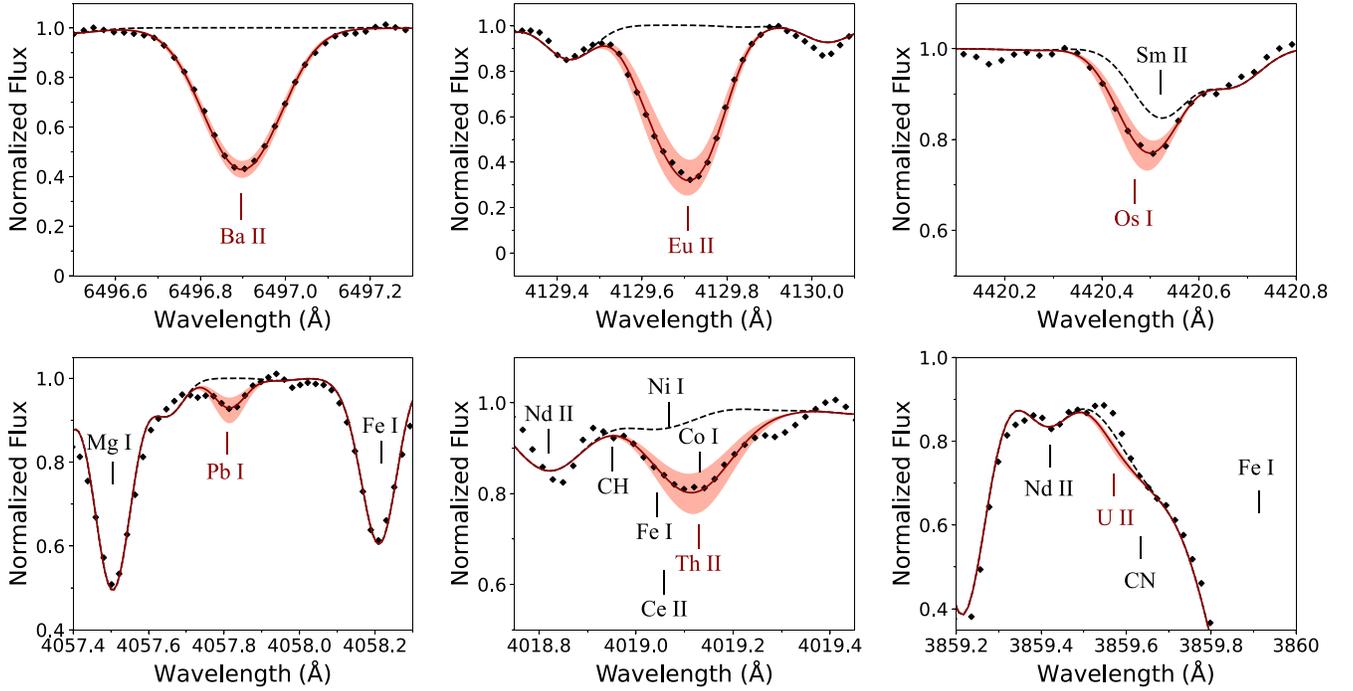

**Figure 1.** The spectral syntheses and the derived abundances for several key elements are presented. The black dots represent the observational spectrum. The red solid lines represent the best-fit syntheses, with an associated uncertainty of 0.2 dex (red shaded regions). The spectral syntheses excluding the elements of interest are shown as black dashed lines.

Compared to the optical spectrum with $R \sim 36{,}000$ (H. Li et al. 2022), the blue-end spectrum exhibits higher resolution ($R \sim 60{,}000$) and S/N. Specifically, the S/N at 4100 Å of the blue-end spectrum (73) is significantly higher than that of the optical spectrum (25). Consequently, the species abundances of J0804+5740 are determined using the blue-end spectrum, with the exception of the O I, Na I, Ca I, Sc II, V I, and Ba II abundances, whose absorption lines are predominantly concentrated in the optical band. The species abundances agree well with those reported in H. Li et al. (2022) within $2\sigma$, except for Co, Zr, and La. The primary reason for the differences in the abundances of Co, Zr, and La is the S/N of the spectrum. Additionally, the blending with Fe I lines and the hyperfine structure (HFS) effect (W. J. Childs & L. S. Goodman 1968; G. H. Guthöhrlein & H. P. Keller 1990; J. C. Pickering 1996) significantly impact the abundance of Co.

### 3.1. Light Neutron-capture Elements

The detection of light neutron-capture elements, particularly those between the first and second peaks, poses significant challenges when using the optical spectrum. We use the blue-end spectrum to derive the abundances of Sr, Y, Zr, Mo, Ru, and Rh. The abundance of Sr is derived through spectral synthesis of two lines at 4077 and 4215 Å. For Y and Zr, which have numerous available lines (14 and 21, respectively) in the blue end of the spectrum, we adopt the abundance values derived from the EWs without any modifications. In the case of Mo, Ru, and Rh, where only a few lines (ranging from one to four) are detectable, we employ spectral synthesis to determine their abundances. It is worth noting that we detect only a faint absorption signature for the Nb II at 3740 Å. This line is significantly affected by spectral noise. Consequently, we could only determine an upper limit for the abundance of Nb.

The deviations in the abundance of light neutron-capture elements in J0804+5740 from the solar $r$-process pattern are observed, particularly for Sr and Zr, as shown in Figure 2. These deviations could be explained by assuming the production of these elements in the "weak" $r$-process (S. Wanajo et al. 2001) and the light-element primary process (C. Travaglio et al. 2004) scenario. Notably, I. U. Roederer et al. (2022) identified good internal consistency in the abundance of light neutron-capture elements among $r$-process-enhanced stars, implying an astrophysical origin independent of the main $r$-process. Furthermore, the fission of transuranic nuclei could have contributed to the abundance of these elements (N. Vassh et al. 2020; I. U. Roederer et al. 2023), as described in Section 4.3.

### 3.2. Heavy Neutron-capture Elements

The abundances for 20 heavy neutron-capture elements are derived, including all the measurable lanthanides, and the upper limit for the abundance of U in J0804+5740 is estimated. The abundances of most elements are obtained through spectral synthesis, taking into account the effects of HFS and isotopic shifts. For Ba, Eu, Yb, and Pb, the solar *r*-process isotope ratios are adopted.

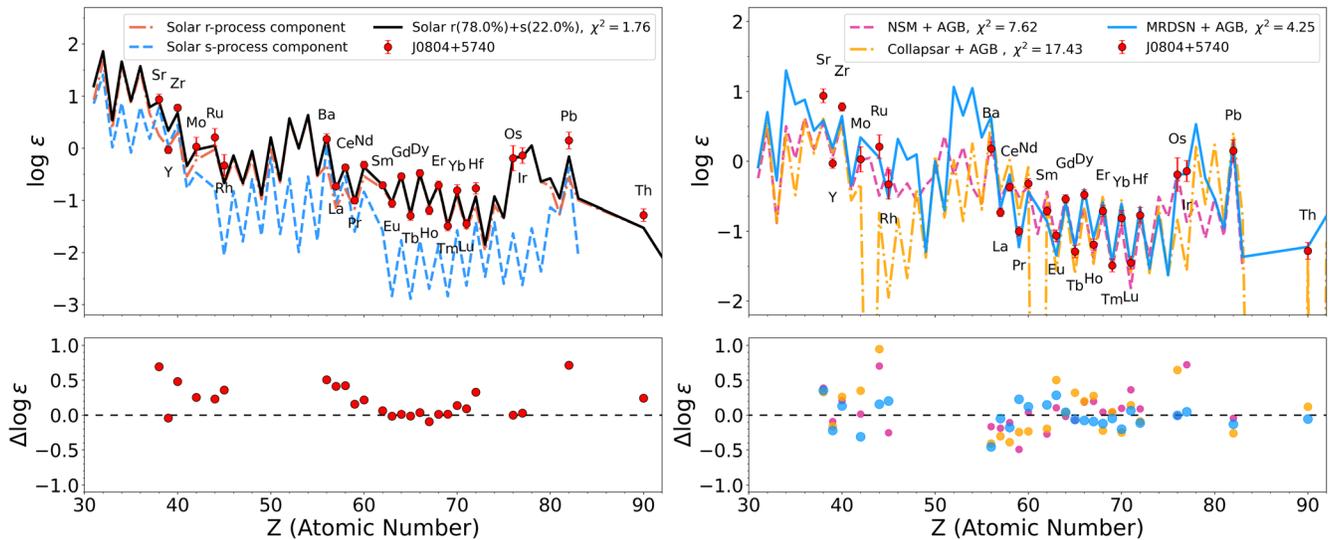

**Figure 2.** Top panel: the heavy-element abundance pattern for J0804+5740, compared with the solar *r*-process and *s*-process patterns (N. Prantzos et al. 2020; top left panel), as well as the theoretical *r*-process models combined with the asymptotic giant branch (AGB) *s*-process model (top right panel). The derived element abundances are depicted with red filled circles, with total error bars. Bottom panel: residuals between the derived element abundances and the solar *r*-process pattern (bottom left panel), as well as the theoretical *r*-process models combined with the AGB *s*-process model (bottom right panel) for J0804+5740.

Like most *r*-process enhanced stars, the abundance pattern of most heavy neutron-capture elements in J0804+5740 agrees well with the solar *r*-process, indicating that the main *r*-process produced these elements in the early Universe (A. Frebel 2018). However, some elements exhibit deviations from the solar *r*-process abundance pattern, particularly Ba, La, and Ce. For instance, [Ba/Eu] = $-0.42 \pm 0.13$ is notably higher than the solar *r*-process value of [Ba/Eu], which varies from $-0.81$ to $-0.90$ (C. Sneden et al. 2008; S. Bisterzo et al. 2014; N. Prantzos et al. 2020), and slightly higher than the value of approximately $-0.6$ observed in most *r*-process-enhanced stars (T. T. Hansen et al. 2018; C. M. Sakari et al. 2018; R. Ezzeddine et al. 2020; E. M. Holmbeck et al. 2020; A. Bandyopadhyay et al. 2024). To account for this, we incorporate solar *s*-process contributions during the fitting process to reproduce the observed abundances of heavy neutron-capture elements. The fitting result, as shown in Figure 2, demonstrates that all lanthanides align well with the solar *r-* + *s*-process.

Os and Ir represent the third *r*-process peak. The abundance of Os is derived from the lines at 4260 and 4420 Å. The Os I line at 4420 Å is significantly blended with Sm, as shown in Figure 1. The abundance of Ir is obtained only from the Ir I 3800 Å. The abundances of both elements are in good agreement with the solar *r*-process.

The abundance of Pb is derived from the Pb I line at 4057 Å, yielding [Pb/Fe] = 0.78. The abundance of Pb is significantly higher than the value of the solar *r*-process. This discrepancy may indicate contamination from the *s*-process. As a third-peak element of the *s*-process, Pb is challenging to measure in *r*-process-enhanced stars. Importantly, Pb is the radioactive decay product of Th and U. As the abundances of Th and U decrease, the Pb abundance will increase by 0.1 dex after 13 Gyr (I. U. Roederer et al. 2024). Due to the nonnegligible contribution of the *s*-process to the abundance of Pb in J0804+5740, we cannot use [Pb/X] as the chronometer pairs to calculate its age.

### 3.3. The Actinides

Th and U are the heaviest elements produced by the *r*-process, yet their detection remains challenging. To date, measurements of Th have been achieved in 53 metal-poor stars, while U detection has been reported in only 8 metal-poor stars (J. J. Cowan et al. 2002; V. Hill et al. 2002, 2017; A. Frebel et al. 2007; V. M. Placco et al. 2017; E. M. Holmbeck et al. 2018; D. Yong et al. 2021; S. P. Shah et al. 2023; I. U. Roederer et al. 2024). In the case of J0804+5740, the Th abundance is derived from the strongest Th II line at 4019 Å. Figure 1 shows the spectral synthesis of this line, demonstrating varying degrees of blending with different species. Notably, this blending effect is significantly reduced in metal-poor stars that lack substantial carbon enhancement ([C/Fe] < 0.7). Unfortunately, due to the significant blending of the U II 3859 Å line with Fe I and CN molecular lines, we can only provide an upper limit for the abundance of U.

Given that $^{232}$Th and $^{238}$U are unstable isotopes with half-lives of 14.05 Gyr and 4.47 Gyr, respectively, we anticipate variations in the ratio of actinides to lanthanides among *r*-process-enhanced stars. Interestingly, approximately one-third of stars with measurable Th abundances exhibit excessive actinide abundances (log $\epsilon$ (Th/Eu) > $-0.35$ or log $\epsilon$ (Th/Dy) > $-0.90$) (L. Mashonkina et al. 2014; E. M. Holmbeck et al. 2019), leading to small or even unrealistic ages calculated from the ratio of actinides to stable lanthanides, if fixed initial abundance ratios are assumed. J0804+5740 is thus classified as an actinide-boost star, with log $\epsilon$ (Th/Eu) = $-0.22$, and is a rare example of actinide boost at relatively high metallicity, as shown in Figure 3. From previous studies, a slight decrease in log $\epsilon$ (Th/Eu) with increasing [Fe/H] could be found: more actinide-boost stars are found at lower metallicities (T. Mishenina et al. 2022; P. Saraf et al. 2023). Our result, however, provides a new data point at relatively high metallicity with high Th/Eu. Further studies of actinide abundances for stars with [Fe/H] > −2.5 are useful to understand the actinide-boost phenomenon.

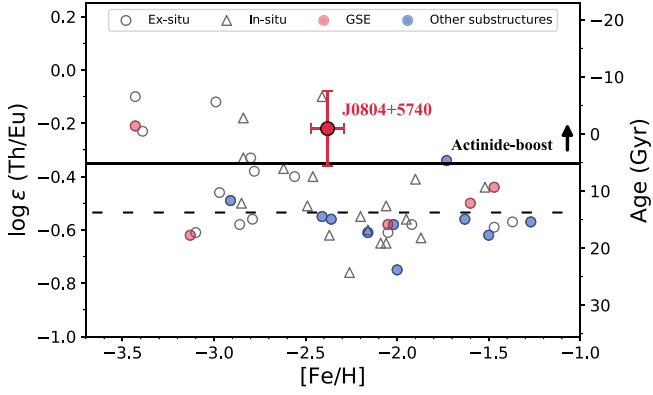

**Figure 3.** $\log \epsilon$ (Th/Eu) as a function of [Fe/H] for *r*-process-enhanced metal-poor stars. Circles represent ex situ stars, while triangles represent in situ stars. The red filled circles represent stars belonging to the GSE, while the blue filled circles denote stars from other substructures, including Wukong, Helmi streams, Thamnos, and Sequoia. The right *Y*-axis indicates the age calculated from $\log \epsilon$ (Th/Eu) (V. Hill et al. 2017). The black solid line corresponds to the actinide-boost criterion proposed by L. Mashonkina et al. (2014). Additionally, the black dashed line represents the age of the Universe (13.772 ± 0.059; C. L. Bennett et al. 2013). The references for the stellar data are compiled from V. M. Placco et al. (2023), supplemented by measurements from S. Honda et al. (2004), D. K. Lai et al. (2008), I. U. Roederer et al. (2009, 2010, 2018, 2024), C. M. Sakari et al. (2018), M. Hanke et al. (2020), D. Yong et al. (2021), M. Gull et al. (2021), P. Saraf et al. (2023), Q. Xing et al. (2024), and A. R. da Silva & R. Smiljanic (2025).

## 4. Discussion

### 4.1. Abundance Pattern

The *s*-process may have made a significant contribution to the abundance patterns of the stars with [Fe/H] > −2.6 (J. Simmerer et al. 2004). As indicated in Figure 2, the surface abundance of J0804+5740 exhibits consistency with the solar *r*- + *s*-process abundance pattern, indicating *s*-process contamination. We use the method as described in R. Jiang et al. (2024) to fit the abundance pattern of J0804+5740. For the abundance of the element *X*, we consider that it arises from both *r*- and *s*-process nucleosynthesis. The calculations were performed using Equation (2):

$$\log \epsilon (X) = \log_{10}(10^{\log \epsilon (X)_r + O_r} + 10^{\log \epsilon (X)_s + O_s}), \quad (2)$$

where $\log \epsilon (X)_r$ and $\log \epsilon (X)_s$ denote the abundance of element *X* from the solar *r*- and *s*-processes, respectively, and $O_r$ and $O_s$ represent their corresponding dilution factors. Based on Equation (2), we find that approximately 78% of the heavy-element abundances on its surface originate from the *r*-process, while only 22% result from the *s*-process. This suggests that the heavy-element composition in J0804+5740 remains predominantly influenced by the *r*-process.

The radial velocity of J0804+5740, measured using a high-resolution spectrum from the Subaru Telescope, is presented in Table 1. Comparison with measurements from Large Sky Area Multi-Object Fiber Spectroscopic Telescope (LAMOST) DR7 (A. L. Luo et al. 2022) and Gaia DR3 shows no significant radial velocity variation within 2σ, indicating that J0804+5740 is unlikely to be part of a binary system. Furthermore, its renormalized unit weight error (RUWE) value of 0.979 suggests that the astrometric observations of J0804+5740 are consistent with a single-star model. Additionally, compared to carbon-enhanced metal-poor (CEMP) stars ([C/Fe] ⩾ 0.7, W. Aoki et al. 2007), J0804+5740 exhibits no significant carbon enrichment ([C/Fe] = 0.15) after accounting for the effect of stellar evolution (V. M. Placco et al. 2014). This suggests that the *s*-process elements in J0804+5740 may not originate from a mass transfer event involving an asymptotic giant branch (AGB) companion star. Instead, it indicates that the birth gas cloud of this star has been contaminated by the *s*-process (M. Gull et al. 2021).

To study the properties of the progenitors that enriched the abundance of J0804+5740, we compare its abundance pattern with different theoretical models. We select three different *r*-process theoretical models: the NSM models (O. Just et al. 2015), the MRDSN models (N. Nishimura et al. 2017), and the collapsar model (Z. He et al. 2024) to determine the origin of *r*-process elements in J0804+5740. Additionally, due to its low metallicity, and the metallicity dependence of the *s*-process, we choose the AGB *s*-process (S. Bisterzo et al. 2010) at low metallicity to replace the solar *s*-process. The final result is shown in Figure 2. The best fit is obtained by combining the MRDSN *r*-process model with $\hat{L}v = 0.2$ and the AGB *s*-process model with [Fe/H] = −2.6 and $M = 1.4 M_\odot$, which reproduces the abundance characteristics of the lanthanides, the third peak of the *r*-process (Os; Ir) and *s*-process (Pb), as well as the actinide-boost feature. However, none of the three models provided a good fit in the region of light neutron-capture elements. By employing the dilution factors of the MRDSNe *r*-process calculated from Equation (2), we determine the dilution mass $M_H = 1.22 \times 10^6 M_\odot$. This result suggests two possible scenarios for the progenitor system of J0804+5740: (1) a dwarf galaxy with a mass of $10^6 M_\odot$ that was accreted by the GSE or (2) a single *r*-process event that enriched $10^6 M_\odot$ of gas within the GSE. Significantly, due to the large nuclear physics uncertainties in *r*-process nucleosynthesis networks, any attempt to compare the abundance patterns between the second and third peaks of the *r*-process with theoretical models remains fraught with risk.

### 4.2. Kinematics

J0804+5740 is classified as a member of the GSE due to its highly radial orbital characteristics. Notably, it is the first actinide-boost star identified within the GSE to date. To further investigate the origins of actinide-boost stars by incorporating kinematic information, we compile a sample consisting of *r*-process-enhanced metal-poor stars with measurable Th abundances from the literature, building upon the discovery of J0804+5740. We use photo-astrometric distances provided by F. Anders et al. (2022) and astrometric data from Gaia DR3 (Gaia Collaboration et al. 2023) to calculate the kinematic and orbital properties of these stars. We only utilized data that exhibit reliable astrometric results (RUWE < 1.4) and relatively precise distance measurements ($\sigma_d/d < 0.4$), ultimately retaining 48 stars in our final sample. We employ the Galactocentric Cartesian coordinate system and assume that the position and velocity of the Sun are $(X, Y, Z)_\odot = (-8.21, 0.0, 0.0208)$ kpc (P. J. McMillan 2017; M. Bennett & J. Bovy 2019) and $(U, V, W)_\odot = (11.1, 12.24, 7.25)$ km s$^{-1}$ (R. Schönrich et al. 2010). The velocity of the local standard of rest ($V_{LSR}$) is adopted as 233.1 km s$^{-1}$. We use the software AGAMA (E. Vasiliev 2019) and the gravitational potential model of the Milky Way from P. J. McMillan (2017) to calculate the dynamical and orbital properties, including the orbital energy ($E$), orbital pericentric radius ($r_{peri}$), orbital apocentric radius ($r_{apo}$), orbital eccentricity (ecc), maximum height from the Galactic plane ($Z_{max}$), and action ($J_r, J_\phi, J_z$). These quantities for J0804+5740 are listed in Table 1.

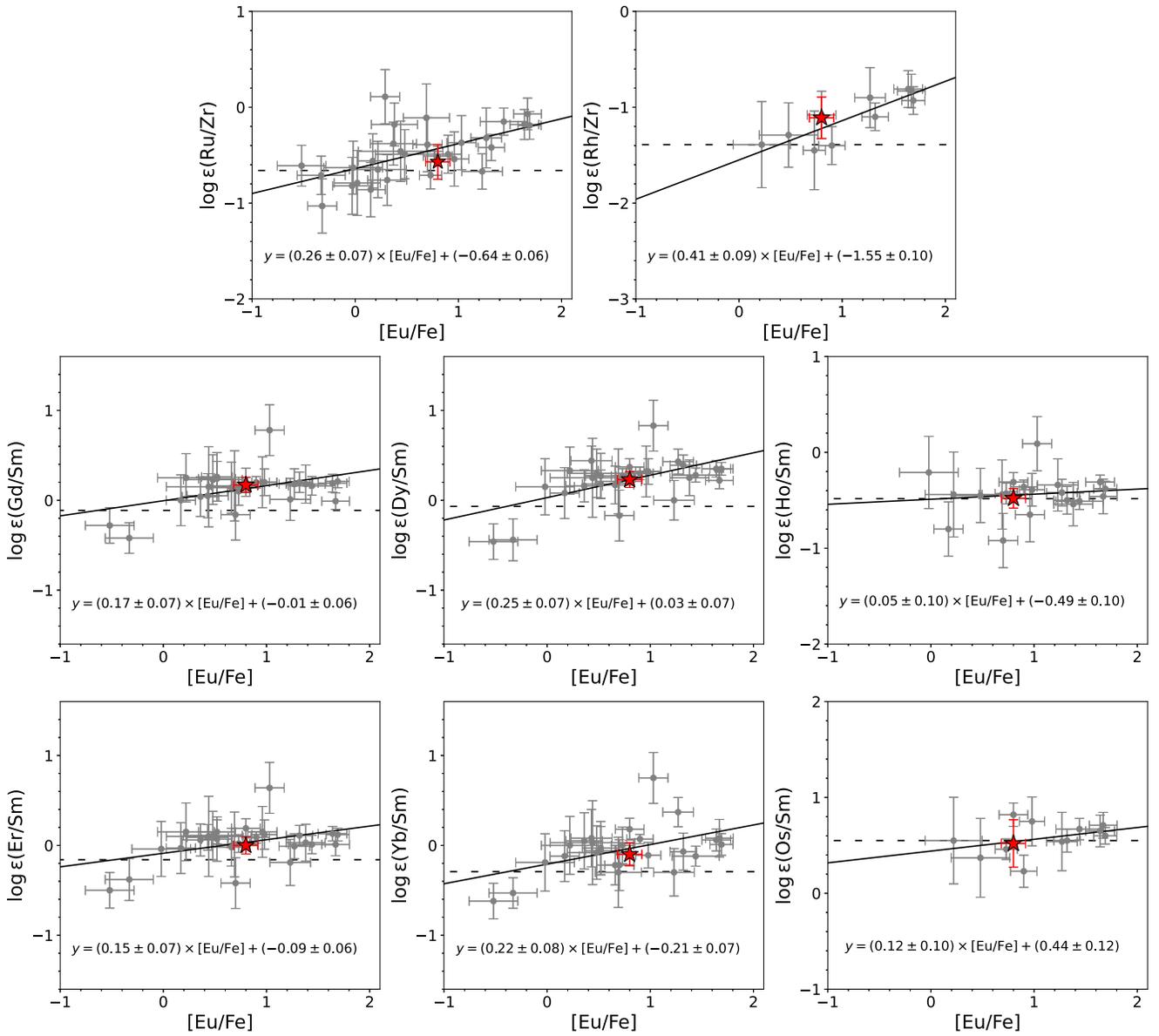

**Figure 4.** The abundance ratio of log $\epsilon$ (X/Zr) for light neutron-capture elements and log $\epsilon$ (X/Sm) for heavy neutron-capture elements as a function of [Eu/Fe]. The gray dots represent the sample stars from I. U. Roederer et al. (2023). J0804+5740 is marked by a red pentagram. The solid black line shows the best fit obtained using the least-squares method (excluding J0804+5740), with the corresponding equation displayed at the bottom of each panel. The dashed black line indicates the baseline abundance calculated from stars with [Eu/Fe] $\leqslant$ 0.3 in I. U. Roederer et al. (2023).

Based on kinematic and orbital parameters, we employ deep-learning methodologies to distinguish between ex situ stars (originating from accreted dwarf galaxies) and in situ stars (formed within the Milky way) in this sample (Z. Li et al. 2024). The classification results are presented in Figure 3. Our analysis reveals that among the nine actinide-boost stars, six are classified as ex situ stars, while only three are categorized as in situ stars. This finding suggests that actinide-boost stars are more likely to originate from dwarf galaxies that were accreted by the Milky Way.

We apply the substructure classification criteria summarized by R. Zhang et al. (2024) and Z. Li et al. (2024) to determine if the sample stars belong to known substructures. Notably, only a small fraction of stars in our sample are associated with known substructures. Specifically, six stars are associated with the GSE, including J0804+5740; three stars are linked to the Wukong; four stars belong to the Helmi streams; two stars are associated with the Thamnos; and one star is linked to Sequoia. Furthermore, another actinide-boost star, CS 30315-0029 (log $\epsilon$ (Th/Eu) = −0.21,C. Siqueira Mello et al. 2014), has also been identified as a member of the GSE. This star is an extremely metal-poor ([Fe/H] = −3.40) r-I star ([Eu/Fe] = 0.68, [Ba/Eu] = −0.51). Importantly, the two actinide-boost stars within the GSE exhibit the highest log $\epsilon$ (Th/Eu) values compared to stars within other substructures. This suggests that the *r*-process events occurring within the GSE are more likely to create conditions conducive to the actinide-boost phenomenon.

The origin of actinide-boost stars remains an unresolved question in astrophysics. Theoretical studies suggest that the actinide-boost phenomenon may arise in environments with significantly low electron fractions ($Y_e = 1/(1 + N_n/N_p)$), such as those associated with NSMs or black hole–neutron-star mergers (E. M. Holmbeck et al. 2019; S. Wanajo et al. 2024). The discovery of J0804+5740 suggests that MRDSNe may also serve as a potential astrophysical site for *r*-process nucleosynthesis, capable of producing the actinide-boost phenomenon (D. Yong et al. 2021). This finding provides critical constraints on *r*-process nucleosynthesis and the production mechanisms of actinide elements in accreted dwarf galaxies such as the GSE.

### 4.3. The Fission of Transuranic Nuclei

Recently, I. U. Roederer et al. (2023) discussed a correlation between the light neutron-capture elements ($44 \leqslant Z \leqslant 47$) and the heavy neutron-capture elements ($63 \leqslant Z \leqslant 78$) in *r*-process enhanced stars. They proposed that the fission fragments of transuranic nuclei enhanced the abundance of these elements. Figure 4 shows the relationship between $\log \epsilon \, (X/Zr)$ and $\log \epsilon \, (X/Sm)$ with [Eu/Fe] for the sample stars from I. U. Roederer et al. (2023). We choose to use the $\log \epsilon \, (X/Sm)$ ratio instead of $\log \epsilon \, (X/Ba)$ because, similar to Ba, the fission fragments of transuranic nuclei do not significantly contribute to the abundance of Sm, while the *s*-process contributes less to the abundance of Sm compared to Ba, as shown in Figure 2. We calculate the baseline abundance of $\log \epsilon \, (X/Sm)$ and its correlation with [Eu/Fe] for their sample stars following the method described by I. U. Roederer et al. (2023). We find that the light and heavy neutron-capture elements in J0804+5740 follow the overall trend, with most element abundance ratios exceeding the baseline ratio. This may indicate potential evidence of fission fragment contributions to the abundances in J0804+5740.

### 5. Conclusion

We present a detailed chemical abundance analysis of J0804+5740, the first actinide-boost star identified in the GSE, characterized by its high actinide-to-lanthanide ratio of $\log \epsilon \, (Th/Eu) = -0.22$. This star was initially discovered through the LAMOST low-resolution survey, and its high-resolution ($R \sim 36{,}000$ and $60{,}000$) spectra were subsequently obtained using Subaru/HDS.

We derive the abundances of 48 species in this star, including Th. The observed abundance pattern aligns well with the solar *r*-process abundance pattern. The MRDSN *r*-process theoretical model with $\hat{L}v = 0.2$ provides a better fit to the abundance features of this star, corresponding to the dilution of *r*-process material into a gas of $10^6 \, M_\odot$. However, the abundance ratios of certain elements suggest an additional small contribution from the *s*-process, indicating a more complex nucleosynthetic origin.

J0804+5740 represents a rare example of an actinide-boost star at relatively high metallicity. Further investigations of actinide-boost stars with [Fe/H] > −2.5 will provide valuable insights into the mechanisms underlying the actinide-boost phenomenon. Furthermore, our comprehensive kinematic analysis reveals that the number of ex situ stars among actinide-boost stars is twice that of in situ stars, suggesting that actinide-boost stars are more likely to originate from accreted dwarf galaxies. Additionally, the discovery of J0804+5740 suggests that MRDSNe may also represent a potential astrophysical site for *r*-process nucleosynthesis, capable of generating conditions that lead to the actinide-boost phenomenon. This finding provides critical constraints on the *r*-process nucleosynthesis mechanisms in accreted dwarf galaxies such as the GSE.

### Acknowledgments

We thank the anonymous referee for valuable and constructive feedback, which has significantly enhanced the quality of the manuscript. This work was supported by the National Key R&D Program of China Nos. 2024YFA1611903 and 2023YFE0107800, the National Natural Science Foundation of China grant No. 12222305, the Strategic Priority Research Program of Chinese Academy of Sciences grant Nos. XDB1160100 and XDB34020205, the International Partnership Program of Chinese Academy of Sciences grant No. 178GJHZ2022040GC, and the science research grants from the China Manned Space Project. This work was also supported by JSPS KAKENHI grant Nos. 20H05855, 21H04499, 22K03688, and 23HP8014.

This research is based on data collected at Subaru Telescope, which is operated by the National Astronomical Observatory of Japan. We are honored and grateful for the opportunity of observing the Universe from Maunakea, which has cultural, historical, and natural significance in Hawaii. Guoshoujing Telescope (the Large Sky Area Multi-Object Fiber Spectroscopic Telescope, LAMOST) is a National Major Scientific Project built by the Chinese Academy of Sciences. Funding for the project has been provided by the National Development and Reform Commission. It is operated and managed by the National Astronomical Observatories, Chinese Academy of Sciences. This work has made use of data from the European Space Agency (ESA) mission Gaia (https://www.cosmos.esa.int/gaia), processed by the Gaia Data Processing and Analysis Consortium (DPAC, https://www.cosmos.esa.int/web/gaia/dpac/consortium). Funding for the DPAC has been provided by national institutions, in particular the institutions participating in the Gaia Multilateral Agreement.

*Facilities:* LAMOST, Subaru.

*Software:* MOOG (C. A. Sneden 1973), TAME (W. Kang & S.-G. Lee 2012), Linemake (V. M. Placco et al. 2021), iSpec (S. Blanco-Cuaresma et al. 2014), Agama (E. Vasiliev 2019), TOPCAT (M. B. Taylor 2005).

### Appendix

Table A1 presents the atomic parameters of the spectral lines used for abundance determination, along with their measured EWs, derived abundances, and the methods employed for abundance derivation.

**Table A1**
Line Atomic Data and Derived Abundances

| Species | Wavelength (Å) | EP (eV) | log gf | EW (mÅ) | Limit Flag | log ε | Method[a] |
|---|---|---|---|---|---|---|---|
| C (CH) | 4322.00 | ... | ... | ... | ... | 5.63 | S |
| N (CN) | 3883.00 | ... | ... | ... | ... | 5.90 | S |
| O I | 6300.30 | 0.00 | −9.78 | 10.90 | ... | 7.15 | S |
| Na I | 5889.95 | 0.00 | 0.11 | 197.50 | ... | 3.98 | S |

**Note.**
[a] The method used to derive the abundance is indicated as follows: "E" denotes that the abundance is derived from EWs, while "S" signifies that the abundance is derived from spectral synthesis.

(This table is available in its entirety in machine-readable form in the online article.)

### ORCID iDs

Yangming Lin ● https://orcid.org/0009-0000-8769-3142
Haining Li ● https://orcid.org/0000-0002-0389-9264
Ruizheng Jiang ● https://orcid.org/0009-0001-0604-072X
Wako Aoki ● https://orcid.org/0000-0002-8975-6829
Satoshi Honda ● https://orcid.org/0000-0001-6653-8741
Zhenyu He ● https://orcid.org/0009-0002-0124-9492
Ruizhi Zhang ● https://orcid.org/0009-0008-1319-1084
Zhuohan Li ● https://orcid.org/0000-0002-1126-9289
Gang Zhao ● https://orcid.org/0000-0002-8980-945X